\documentclass[prd,a4paper,preprint,preprintnumbers,superscriptaddress,nofootinbib]{revtex4-2}

\usepackage[a4paper, hdivide={1.91cm,,1.165cm}, vdivide={1.83cm,,3.0cm}]{geometry}

\usepackage{amsmath}
\usepackage{graphicx}
\usepackage{xspace}
\usepackage{color}
\usepackage{units}
\usepackage{slashed}
\usepackage[hyperfootnotes=false]{hyperref}
\hypersetup{linkcolor=red,
citecolor=green,
filecolor=cyan,
urlcolor=magenta}
\usepackage{gensymb}
\usepackage{booktabs}
\usepackage{multirow}
\usepackage{xcolor}
\usepackage{setspace}

\newcommand{\beq}{\begin{eqnarray}}
\newcommand{\eeq}{\end{eqnarray}}
\def\mcdot{\!\cdot\!}
\def\ds{\displaystyle}
\def\beq{\begin{equation}}
\def\eeq{\end{equation}}
\def\bea{\begin{eqnarray}}
\def\eea{\end{eqnarray}}
\def\ve{\vert}
\def\nnb{\nonumber}
\def\nnb{\nonumber}

\def\es{ &=& }
\def\ar{&+& }
\def\ek{&-& }
\def\cp{&\times&}

\begin{document}


\title{ Weak $\Xi_{QQ} \to \Sigma_Q \ell^+ \ell^-$ decays induced by FCNC in QCD}

\author{T.~M.~Aliev}
\email{taliev@metu.edu.tr}
\affiliation{Department of Physics, Middle East Technical University, Ankara, 06800, Turkey}

\author{S.~Bilmis}
\email{sbilmis@metu.edu.tr}
\affiliation{Department of Physics, Middle East Technical University, Ankara, 06800, Turkey}
\affiliation{TUBITAK ULAKBIM, Ankara, 06510, Turkey}

\author{M.~Savci}
\email{savci@metu.edu.tr}
\affiliation{Department of Physics, Middle East Technical University, Ankara, 06800, Turkey}

\date{\today}

\begin{abstract}
With the discovery of the doubly heavy $\Xi_{cc}$ baryon, comprehensive studies of the properties of the doubly heavy baryons are started. In the present work, we examine the $\Xi_{bb} \to \Sigma_b \ell^+ \ell^-$ and $\Xi_{cc} \to \Sigma_c \ell^+ \ell^-$ decays induced by flavor-changing neutral currents (FCNC) in the framework of the light-cone sum rules. After obtaining the sum rules for the form factors induced by the tensor current, the branching ratios of the relevant transitions are estimated. We found that the branching ratio for the $c \to u$ transition is around five orders smaller than the $b \to d$ transition. Our findings are also compared with other approaches. 
\end{abstract}

\maketitle

\newpage


\section{Introduction\label{intro}}
The quark model has been quite successful in the classification of the hadrons. However, up to now, only the hadron state $\Xi_{cc}^{++}$ has been discovered among all the baryons containing double heavy quarks anticipated by the quark model~\cite{LHCb:2017iph,LHCb:2018pcs,LHCb:2018zpl}. The detailed analysis to determine the properties of these hadrons is crucial to precisely testing the Standard Model(SM) as well as looking for new physics effects. Weak decays induced by flavor-changing neutral current (FCNC) of doubly heavy baryons are an ideal framework to check SM predictions at the loop level. The new physics effects can manifest themselves in these interactions either by modifying the so-called Wilson coefficients existing in the SM without introducing new operators or by introducing new effective operators.

The observation of the doubly heavy hadrons triggered many theoretical studies on this subject (see \cite{Aliyev:2022rrf} and the references therein). In this context, a comprehensive analysis of the weak decays of doubly heavy baryons occupies a special place.  The main ingredient of weak decays is the transition matrix elements between the initial and final states due to the weak currents of quarks. These matrix elements are parametrized in terms of the form factors. Calculation of the form factors is the main ingredient of studying the weak decays which belong to the non-perturbative domain of the QCD. For this reason, some non-pertubative methods are needed for their calculation. Among various non-perturbative methods, sum rules method that is based on the fundamental QCD Lagrangian occupies an exceptional place. Form factors of some of the doubly heavy baryons due to the charged current are already studied with the traditional and light cone version of the sum rules in \cite{Shi:2019hbf}, and \cite{Shi:2019fph,Hu:2019bqj}, respectively. It should be noted here that form factors
of the doubly heavy baryons decaying to single heavy baryons are studied in the works \cite{Wang:2017mqp,Zhao:2018mrg,Xing:2018lre,Ke:2019lcf} in the framework of the light-front quark model. Moreover, FCNC processes of the doubly heavy baryons are studied within this approach in \cite{Hu:2020mxk} and \cite{Xing:2018lre}. It should also be noted that the FCNC-induced decay of $\Xi_{QQ} \to \Lambda_Q \ell^+ \ell^-$ decay within the light-cone sum rules are studied in ~\cite{Aliev:2022maw}.

Before we delve into our analysis, we would like to say a few words about the SU(3) classification of the doubly heavy baryons.
A doubly heavy baryon contains two heavy and one light quark. Doubly heavy baryons with  $J^P = {1 \over 2}^+$ in the $cc$ sector
are $\Xi_{cc}^{++}$, $\Xi_{cc}^+$ and $\Omega_{cc}^+$, and those in the $bb$ sector are  $\Xi_{bb}^0$, $\Xi_{bb}^-$ and $\Omega_{bb}^-$. Additionally, there are two sets of baryons in the $bc$ sector which are antisymmetric or symmetric under the interchange of $b$ and $c$ quarks. While the single heavy baryons $\Lambda_Q$, $\Xi_Q$ belong to the triplet(anti) representation, $\Sigma_Q, \Sigma_{Q^\prime}$, and $\Omega_Q$ baryons lie in the sextet representation of SU(3). 

In the present work, we study the $\Xi_{cc}^+ \to \Sigma_c^+ \ell^+ \ell^-$, and $\Xi_{bb}^{0} \to \Sigma_b^{0} \ell^+ \ell^-$ decay in the framework of the light cone QCD sum rules method (LCSR). This method is an extension of the traditional QCD sum rules method \cite{Shifman:1978bx}, and to the light cone \cite{Chernyak:1983ej,Balitsky:1989ry}. Within the framework of this method, many aspects of the hadron physics are studied (see the review
\cite{Colangelo:2000dp}). In the framework of the LCSR method, instead of the local operator product expansion (OPE), the light cone expansion of the non-local operators is used. Moreover, in this method, light-cone distribution amplitudes appear instead of the local condensates, and OPE is performed over twists rather than the dimensions of the local operators.

The paper is organized as follows. In Section~\ref{sec:2}, we derive the sum rules for the relevant form factors for the $\Sigma_{QQ} \to \Sigma_Q \ell^+ \ell^-$ decay in framework of the LCSR method. Numerical
analysis of these form factors is presented in Section~\ref{sec:3}. In this section, we also estimate the branching ratios of the corresponding decay using these form factors.
Conclusions and discussions of the obtained results are presented in the last Section~\ref{sec:conclusion}.
\section{Sum rules of the transition form factors for the $\Xi_{QQ} \to \Sigma_Q \ell^+ \ell^-$ decays}
\label{sec:2}
The flavor-changing neutral $b \to q (d \text{ or } s) \ell^+ \ell^- $ transitions up to mass dimension six is described by the standard weak effective field theory~\cite{Buras:1994dj}. The effective Hamiltonian for this transition can be written as
\begin{equation}
  \label{eq:1}
  H_{eff} = - \frac{2 G_F}{\sqrt{2}} \frac{\alpha}{4 \pi} V_{tb} V_{tq}^* \sum C_i^{eff}(\mu) \mathcal{O}_i
\end{equation}
where $G_F$ is the Fermi constant, $V_{ij}$ are the elements of Cabibbo-Kobayashi-Maskawa (CKM) matrix elements, $\alpha$ is the electromagnetic coupling, $C_i^{eff}(\mu)$ are the short-distance Wilson coefficients and $\mathcal{O}_i$ are the local effective field operators.

For the decays under consideration, only operators $\mathcal{O}_7$, $\mathcal{O}_9, \mathcal{O}_{10}$ 
\begin{equation}
  \label{eq:2}
  \begin{split}
    \mathcal{O}_7 &= m_b \bar{q} \sigma_{\mu \nu} (1+\gamma_5) b F_{\mu \nu} \\
    \mathcal{O}_9 &= \bar{q} \gamma_\mu (1 - \gamma_5) b \bar{l} \gamma^\mu l \\
    \mathcal{O}_{10} &=  \bar{q} \gamma_\mu (1-\gamma_5) b \bar{l} \gamma^\mu \gamma_5 l \\
  \end{split}
\end{equation}
are significant at the scale $\mu = m_Q$. It should be noted that the four-quark operators induced by the W-boson exchange (or penguin annihilation) can also contribute to the considered transition. However, these contributions have not been estimated systematically. So-called ``charm-loop effects'' studied for B-meson decays~\cite{Khodjamirian:2010vf}. These effects might also be important for baryon counterparts. These contributions should be accurately calculated for precise determination of the form factors. However, the effects of these contributions are beyond the scope of this work.

At the quark level, the $\Xi_{QQ} \to \Sigma_Q \ell^+ \ell^-$ decays take
place through the $c \to u$ or $b \to d$ transitions. The hadronic matrix
elements for the $\Xi_{cc} \to \Sigma_c \ell^+ \ell^-$ and 
$\Xi_{bb} \to \Sigma_b \ell^+ \ell^-$ decays are determined by sandwiching
the transition currents between the initial and final hadron states. For example, the matrix element for the $ b \to q \ell^+ \ell^-$
transition amplitude between the initial and final hadron states can be
written 
\bea
\label{edb01}
{\cal M} \es {G_{F}\alpha \over 2 \sqrt{2} \pi} V_{tb} V_{tq}^* \Bigg[ \big(
C_9^{eff} \langle \Sigma_b | \bar{q} \gamma_\mu (1-\gamma_5) b | \Xi_{bb}
\rangle - {2 m_b \over q^2} C_7^{eff} \langle \Sigma_b | \bar{q}
i\sigma_{\mu\nu} q^\nu (1+\gamma_5) b | \Xi_{bb} \rangle \Big) \bar{l} \gamma^\mu l \nnb \\
\ar C_{10}^{eff} \langle \Sigma_b | \bar{q} \gamma_\mu (1-\gamma_5) b 
| \Xi_{bb} \rangle \bar{l} \gamma^\mu \gamma_5 l \Bigg]~.
\eea
The effective Wilson coefficient $C_9^{eff}$ for the $b \to q \ell^+ \ell^-$
transition is,
\bea
\label{edb05}
C_9^{eff} \es \Bigg\{ C_9 +  h(m_c,q^2) C_0 +
\lambda_u \Big[  h(m_c,q^2) -  h(m_u,q^2) \Big] (3 C_1 + C_2) \nnb \\
\ek {1 \over 2} h(m_b,q^2) (4 C_3 + 4 C_4 + 3 C_5 + C_6) -
{1 \over 2} h(0,q^2) (C_3 + 3 C_4) \nnb \\
\ar {2\over 9}  (3 C_3 + C_4 + 3 C_5 + C_6)\Bigg\}~,
\eea
where,
\bea
\label{nolabel}
\lambda_u \es {V_{ud}^\ast V_{ub} \over  V_{tb}^\ast V_{tq}}~,\mbox{and,~~}
C_0 = 3 C_1 + C_2 + 3 C_3 + C_4 + 3 C_5 + C_6 ~,
\eea
\bea
\label{edb03}
h(m_q,q^2) \es -{8\over 9} \ln {m_q\over m_c} + {8 \over 27} + {4 \over 9} x
- {2 \over 9} (2+x) \sqrt{\ve 1-x \ve} \nnb \\
\cp \Bigg[ \Theta(1 - x_q) \Bigg( \ln {1  +
\sqrt{1 - x_q} \over {1  -  \sqrt{1 - x_q}}} - i \pi \Bigg)
+ \Theta(x_q - 1) \, 2 \, \arctan {1 \over \sqrt{x_q - 1}} \Bigg]~,\\ \nnb \\
h(0,q^2) \es {8 \over 27} - {4 \over 9} \ln{q^2 \over m_c^2} +
{4\over 9} i \pi~.
\eea
In the last equation, $x_q = {4m_q^2/q^2}$ and $\Theta(x)$ is the Heaviside step function. Both $h(m_c,q^2)$, $h(m_b,q^2)$ can be obtained from $h(m_q,q^2)$ by making the replacements $m_q \to m_c$ and $m_q \to m_b$, respectively. The numerical
values of the Wilson coefficients  $C_7^{eff}$ and $C_{10}^{eff}$ as well as the other $C_i$ for the $b \to d$ transition can be
found in \cite{Altmannshofer:2008dz}.

The matrix element for the $\Xi_{cc} \to \Sigma_c \ell^{+} \ell^{-}$ can be obtained from Eq.~\eqref{edb01}
with the help of the following replacements:
\begin{equation}
  \label{eq:3}
  V_{tb} V_{tq}^* \to V_{cd} V_{ud}^* + V_{cs} V_{bs}^* \hspace{1cm} m_b \to m_c \hspace{1cm} \Xi_{bb} \to \Xi_{cc}, 
\end{equation}
and replace $C_i^{\text{eff}}$ for the b-quark case with the corresponding c-quark counterparts given as below.

The effective Wilson coefficient $C_9^{eff}$ for $c \to u$ transition is
given as \cite{deBoer:2015boa},
\bea
\label{edb02}
C_9^{eff} \es 
C_9 + h_c(m_c,q^2) \Bigg(7 C_3 +{4\over 3} C_4 + 76 C_5 + 
{64\over 3} C_6 \Bigg) \nnb \\
\ek
h_c(m_c,q^2) (3 C_3 + 30 C_5) + 
{4\over 3} h(0,q^2) \Bigg(3 C_3 + C_4 + {69\over2} C_5 + 16 C_6 \Bigg) +
{8\over 3} (C_3 + 10 C_5)  \nnb \\
\ek \Big[V_{cd}^\ast V_{ud} h(0,q^2) + V_{cs}^\ast V_{us} h(m_q,q^2) \Big] 
\Bigg( {2\over 3} C_1 + {1\over 2} C_2 \Bigg)~,
\eea
The values of the Wilson coefficients are presented in \cite{deBoer:2016dcg}, and the effective Wilson coefficient $C_7^{eff}$
is given in \cite{deBoer:2015boa}, which we shall use in further numerical analysis.
Note also that, due to the GIM cancellation, $C_{10}^{eff}$ is zero.

It should be noted here that, $C_9^{eff}$, which appears in the $c \to u$ and
$b \to d$ transitions, receives contributions also from vector mesons (long-distance effects). Long-distance contributions are only significant when $q^2$ is close to the mass of the corresponding vector mesons. However, far from these points, these effects are small; hence, we take only short-distance effects into account.  After these preliminary remarks, we now proceed to calculate the form factors in the framework of the QCD sum rules.

The matrix elements entering into Eq.\eqref{edb01} are parametrized
in terms of the form factors in the following way,
\bea
\label{edb06}
&&\langle \Sigma_Q(p) | \bar{q} \gamma_\mu (1 - \gamma_5) Q |\Xi_{QQ}(p+q)
\rangle =
\bar{u}_{\Sigma_Q}(p) \Bigg[\gamma_\mu f_1(q^2) + {i\sigma_{\mu\nu} q^\nu  \over m_{\Xi_{QQ}}}
f_2(q^2) \nnb \\
\ar {q_\mu  \over m_{\Xi_{QQ}}} f_3(q^2)
- \gamma_\mu \gamma_5 g_1(q^2) - {i\sigma_{\mu\nu} \gamma_5 q^\nu  \over m_{\Xi_{QQ}}}
g_2(q^2) - {\gamma_5 q_\mu  \over m_{\Xi_{QQ}}} g_3(q^2) \Bigg]
u_{\Xi_{QQ}}(p+q)~, \\ \nnb \\
\label{edb07}
&&\langle \Sigma_Q(p) | \bar{q} i \sigma_{\mu\nu} q^\nu(1 + \gamma_5) Q | \Xi_{QQ}(p+q)
\rangle = 
\bar{u}_{\Sigma_Q}(p) \Bigg[(\gamma_\mu q^2 - q_\mu \slashed{q}) {f_1^T(q^2) \over
m_{\Xi_{QQ}}} \nnb \\
\ar i\sigma_{\mu\nu} q^\nu f_2^T(q^2)
+ (\gamma_\mu q^2 - q_\mu \slashed{q}) \gamma_5 {g_1^T(q^2) \over  m_{\Xi_{QQ}}} +
  i\sigma_{\mu\nu} \gamma_5 q^\nu g_2^T(q^2) \Bigg] u_{\Xi_{QQ}}(p+q)~.
  \eea
  where $u_{\Sigma_Q}$ and $u_{\Xi_{QQ}}$ are the spinors of the single and doubly heavy baryons.
The form factors $f_i$ and $g_i$ are estimated in the framework of the light
cone sum rules method in \cite{Hu:2019bqj}, and for this reason, we pay attention to the calculation of the form factors $f_i^T$ and $g_i^T$ using
the LCSR method only.  

In order to calculate the form factors $f_i^T$ and $g_i^T$ in the framework of
the LCSR method, we start with the following form correlation function.

\bea
\label{edb08} 
 \Pi_\mu (p,q) \es i \int d^4x e^{iq \cdot x} \langle \Sigma_Q (p) | T \big\{ \bar{q}
i \sigma_{\mu \nu} q^\nu (1+\gamma_5) Q \; \bar{j}_{\Xi_{QQ}} \big\} | 0 \rangle ~,
\eea
where $Q=c \text{ or }b$. The interpolating current of the $\Xi_{QQ}$ baryon is,
\bea
\label{edb09}
j_{\Xi_{QQ}} (0) \es \varepsilon^{abc} (Q^{aT} C \gamma^\mu Q^{b})
\gamma_\mu \gamma_5 q^c~,
\eea
where $a,b,c$ are the color indices. In the LCSR method, the expression 
of the correlation function is obtained in two different ways. One of the
representations can be written in terms of the hadrons, and the other is from the QCD side, i.e. in terms of the quarks and gluons.
On the hadronic side, the correlation function is obtained by inserting
hadronic states with the quantum numbers of $\Xi_{QQ}$ baryon. Then, isolating the ground state contributions of the $\Xi_{QQ}$ baryon, we get,
\bea
\label{edb10}
\Pi_\mu (p,q) \es
{\langle \Sigma_Q (p) | \bar{q}
i \sigma_{\mu\nu} q^\nu (1+\gamma_5) Q | \Xi_{QQ}(p+q) \rangle \langle \Xi_{QQ}(p+q)
\ve \bar{j}_{\Xi_{QQ}} (0) \rangle \over m_{\Xi_{QQ}}^2 - (p+q)^2}~.
\eea 
The second matrix element can be written as,
\bea
\label{nolabel}
\langle \Xi_{QQ}(p+q) \ve \bar{j}_{\Xi_{QQ}} (0) \rangle \es f_{\Xi_{QQ}}
\bar{u}_{\Xi_{QQ}}(p+q)~,\nnb
\eea
and the first matrix element which is defined in terms of the form factors,
is given in Eq.\eqref{edb07}.

Using the completeness condition of the Dirac bispinors, we get the following result for the correlation function in terms of hadrons
\bea
\label{edb11}
\Pi_\mu (p,q) \es { f_{\Xi_{QQ}} \over m_{\Xi_{QQ}}^2 - (p+q)^2} \bar{u}(p)
\Bigg\{{f_1^T \over m_{\Xi_{QQ}} } (\gamma_\mu q^2 - q_\mu \slashed{q}) +
i f_2^T \sigma_{\mu\nu} q^\nu \nnb \\
\ar {g_1^T \over m_{\Xi_{QQ}} } (\gamma_\mu q^2 - q_\mu \slashed{q}) \gamma_5 +
i g_2^T \sigma_{\mu\nu} q^\nu \gamma_5 \Bigg\} (\slashed{p} + \slashed{q} +
m_{\Xi_{QQ}})\\
\es { f \over m_1^2-(p+q)^2} \bar{u}(v)
\Bigg\{ \Bigg[\Bigg( {m_2 \over m_1} - 1 \Bigg) f_1^T + f_2^T \Bigg] \slashed{q} q_\mu
+ 2 m_2 f_2^T \slashed{q} v_\mu \nnb \\
\ar \Bigg[\Bigg( {m_2 \over m_1} + 1 \Bigg) g_1^T + g_2^T \Bigg] \slashed{q}
\gamma_5 q_\mu - 2 m_2 g_2^T \slashed{q} \gamma_5 v_\mu\Bigg\} +
\mbox{other structures.} \nnb
\eea
In the last step of the derivation, we used the heavy quark limit, i.e., $p_\mu \to m_{\Sigma_Q} v_\mu$, 
and for brevity we replaced $m_{\Xi_{QQ}}$ by $m_1$, $m_{\Sigma_Q}$ by $m_2$, and
$f_{\Xi_{QQ}}$ by $f$. Having obtained the expression of the correlation function from the hadronic side, let us turn our attention to the calculation of the correlation function from the QCD side.

After using the Wick theorem for the correlation function from the QCD side, we get
\bea
\label{edb12}
\Pi_\mu (p,q) \es
\int d^4x e^{i(q\cdot x)} \Big\{ [i\sigma_{\mu\nu}q^\nu
(1+\gamma_5)]_{\alpha\beta} (\gamma_5 \gamma_\tau)_{\rho\gamma} 
(C\gamma_\tau)_{\sigma\phi} \epsilon^{abc} \nnb \\
\cp\langle \Sigma_Q (p) | \left(-\bar{q}_\alpha^e(x) \bar{q}_\rho^c(0)
\bar{Q}_\phi^a (0) S_{\beta\sigma}^{eb} (x) +
\bar{q}_\alpha^e(x) \bar{q}_\rho^c(0)
\bar{Q}_\phi^b(0) S_{\beta\phi}^{ea} (x)\right)| 0 \rangle \Bigg\}~.
\eea 

The matrix element $\varepsilon^{abc} \langle \Sigma_c (p) |
\bar{q}_\alpha^a (t_1) q_\beta^b (t_2)  \bar{Q}_\rho^c (0) | 0 \rangle$
appearing in Eq.\eqref{edb12} is determined in terms of the heavy baryon
DAs. The light cone distribution amplitudes are studied in~\cite{Ali:2012pn}.
In determining the parameters appearing in DAs, the standard sum rules
method in heavy quark mass limit is considered. The distribution amplitudes
of the $\Sigma_Q$ baryon in the sextet representation of SU(3) are determined in the
following way,
\bea
\label{nolabel}  
{\bar{v}^\mu \over v_+} \varepsilon^{abc} \langle 0 | q_1^{aT} C \slashed{n}
q_2^b (t_2) h_\gamma^c (0) | \Sigma_Q (v) \rangle \es
{1\over \sqrt{3}} \psi_2 (t_1,t_2) f^{(1)} \varepsilon_\parallel^\mu
u_\gamma ~, \nnb \\
i {\bar{v}^\mu \over v_+} \varepsilon^{abc} \langle 0 | q_1^{aT} C
\sigma_{\alpha\beta} q_2^b (t_2) h_\gamma^c (0) | \Sigma_Q (v) \rangle \es
{1\over \sqrt{3}} \psi_3^a (t_1,t_2) f^{(2)} \varepsilon_\parallel^\mu
u_\gamma ~, \nnb \\
\bar{v}^\mu \varepsilon^{abc} \langle 0 | q_1^{aT} C 
q_2^b (t_2) h_\gamma^c (0) | \Sigma_Q (v) \rangle \es
{1\over \sqrt{3}} \psi_3^\sigma (t_1,t_2) f^{(2)} \varepsilon_\parallel^\mu
u_\gamma ~, \nnb \\
- v_+ \bar{v}^\mu \varepsilon^{abc} \langle 0 | q_1^{aT} C \bar{\slashed{n}}
q_2^b (t_2) h_\gamma^c (0) | \Sigma_Q (v) \rangle \es
{1\over \sqrt{3}} \psi_4 (t_1,t_2) f^{(1)} \varepsilon_\parallel^\mu
u_\gamma ~, \nnb
\eea
where $\psi_i$ are the distribution amplitudes with definite twist, $t_i$
are the distance between the $i$th light quark and the origin along the
direction of $n$, $n^\mu$ and $\bar{n}^\mu$ are the two light vectors,
$\bar{v}^\mu = {\ds{{1\over2}\left({n^\mu \over v_+} - v_+ \bar{n}^\mu \right)}}$,
$v^\mu = {\ds{{1\over2}\left({n^\mu \over v_+} + v_+ \bar{n}^\mu \right)}}$,
and the space coordinates are taken as $t_i n^\mu$.
In further discussion we will work in the rest frame  of the $\Sigma_Q$
heavy baryons, i.e., $v_+=1$. Here, $u_\gamma(v)$ is the heavy baryon spinor and $h_\gamma$ is the static heavy quark field in HQET.

Here we would like to make the following remark. The light cone DAs of heavy baryons are obtained in the HQET in terms of four-velocity and heavy quark field, $h$. However, in QCD, heavy baryon state, $|\Sigma \rangle$, is described by the momentum $p$ and heavy quark field by $Q$ (see Eq.~\ref{edb12}). Therefore, these quantities should be transformed to the HQET counterparts. The heavy quark field, $Q$, should be replaced by the corresponding heavy quark effective field $h(0)$, i.e. $Q(0) \rightarrow h(0)$. In addition, the heavy baryon state can be written in terms of the HQET baryon state by using $|\Sigma_Q(p) \rangle = \sqrt{m_2} | \Sigma(v) \rangle + \mathcal{O}(1/m_2)$. In HQET, since the higher order of the inverse heavy quark mass terms can be neglected, we obtain $\Sigma_Q(p) \rangle = \sqrt{m_2} | \Sigma(v) \rangle$. Applying this transformation to both sides of the correlation function, we see that the replacement $|\Sigma(p) \rangle \rightarrow | \Sigma(v) \rangle$ can be made safely. However, this transformation is only valid for tree-level calculations. When $\mathcal{O}(\alpha_s)$ corrections are taken into account, the matching relations among the QCD currents and HQET currents should be used (see ~\cite{Grinstein:2004vb}). In this work, we neglected the NLO corrections. 

As a result, the matrix element $\epsilon^{abc} \langle \Sigma_Q (v) | q_{1 \alpha}^{a} (t_1) \bar{q}_{2 \beta}^b(t_2) h_\gamma^c(0) | 0 \rangle$ in terms of $\Sigma_Q$ distribution amplitudes can be written as
\bea
\label{edb13}
\varepsilon^{abc} \langle \Sigma_Q (v) | \bar{q}_{1\alpha}^a(t_1) \bar{q}_{2\beta}^b(t_2)
\bar{h}_\gamma^c(0) | 0 \rangle = {\sum_{i=1}^4} A_i (\bar{u}_\Sigma \slashed{\bar{v}}
\gamma_5)_\gamma (C^{-1} \Gamma_i)_{\alpha\beta}~,
\eea
where

\begin{align}
  \label{edb14}
  A_1 & = {f^{(1)} \over 8} \psi_2 (t_1,t_2)~,             & A_3 &= { f^{(2)} \over 4} \psi_3^{(s)} (t_1,t_2)~, \nnb \\
  A_2 & = -{ f^{(2)} \over 8} \psi_3^{(\sigma)} (t_1,t_2)~,  & A_4 &= { f^{(1)} \over 8} \psi_4 (t_1,t_2)~, 
\end{align}
and $\Gamma_1 = \bar{\slashed{n}}$, $\Gamma_2 = i\sigma_{\alpha\beta} n^\alpha
\bar{n}^\beta$, $\Gamma_3 = I$, $\Gamma_4 = \slashed{n}$. The distribution
amplitudes $\psi$ are defined as,
\bea
\label{nolabel}
\psi(t_1,t_2) \es \int_0^\infty dw w \int_0^1 du e^{-iw(t_1 u + t_2\bar{u})}
\psi(t_1,t_2)~,\nnb
\eea
where $\bar{u}=1-u $, $t_i = v x_i$, $w_2 = \bar{u} w$ and $w$ is the total light diquark
momentum. Although the DAs are presented only for the bottomed baryons in
\cite{Ali:2012pn,Bell:2013tfa}, one can use these DAs for the baryons containing charm quarks as well in the
heavy quark mass limit. In the present work, both $\Sigma_c$ and $\Sigma_b$
are described by the same DAs given in \cite{Ali:2012pn}. Their explicit forms
are,
\bea
\label{edb15}  
\psi_2(u,w) \es w^2 \bar{u}u \sum_{n=0}^2 {a_n\over \varepsilon_n^4}
{C_n^{3/2} (2 u -1) \over | C_n^{3/2} |^2 } e^{-w/\varepsilon_n}~, \nnb \\
\psi_4(u,w) \es w^2 \bar{u}u \sum_{n=0}^2 {a_n\over \varepsilon_n^2}
{C_n^{1/2} (2 u -1) \over | C_n^{1/2} |^2 } e^{-w/\varepsilon_n}~, \nnb \\
\psi_3^{(\sigma,s)}(u,w) \es {w \over 2} \bar{u}u \sum_{n=0}^2 {a_n\over
\varepsilon_n^3}
{C_n^{1/2} (2 u -1) \over | C_n^{1/2} |^2 } e^{-w/\varepsilon_n}~.
\eea
The values of the parameters $a_0,~a_1,~a_2$, and
$\varepsilon_0,~\varepsilon_1,~\varepsilon_2$ are given in \cite{Ali:2012pn},
and $C_n^{\lambda}(2 u-1)$ is the Gegenbauer polynomial. 

Substituting the light-cone distribution amplitudes Eq\eqref{edb13} into Eq\eqref{edb12}
and using the heavy quark propagator in momentum 
representation and performing integration over $x$, the correlation function
at the QCD level can be written as,
\bea
\label{edb16}
\Pi_\mu^{QCD} [(p+q)^2,q^2] \es
\int du \int dw  {\sum_{n=1}^3} \Bigg\{
 {\rho_n^{(1)} (u,w) \over (\Delta - m_Q^2)^n} q_\mu \slashed{q} +
 {\rho_n^{(2)}(u,w) \over (\Delta - m_Q^2)^n} \slashed{q} v_\mu \\ \nnb
 \ar  {\rho_n^{(3)}(u,w) \over (\Delta - m_Q^2)^n} q_\mu \slashed{q} \gamma_5 
 +  {\rho_n^{(4)}(u,w) \over (\Delta - m_Q^2)^n} \slashed{q} v_\mu \gamma_5 + \text{other structures}
 \Bigg\}~,
\eea
where the invariant functions $\rho_n^{(i)},~(i=1,2,3,4)$ are presented in Appendix~\ref{app:formulas}, and 
\bea
\Delta = \frac{\bar{u} w}{m_{\Sigma_Q}} (p+q)^2 + q^2 ( 1- \frac{\bar{u} w}{m_{\Sigma_Q}}) - 
\bar{u} w (m_\Sigma - \bar{u}w)~.
\eea 

Matching the coefficients of the structures $\slashed{q} q_\mu$, $\slashed{q} v_\mu$, $\slashed{q}
\gamma_5 q_\mu$, and $\slashed{q} \gamma_5 v_\mu$ in 
both representations of the correlation function and applying the Borel
transformation with respect to the variable $-(p+q)^2$ in order to enhance
the contributions of the ground states, and suppress the higher states and
continuum contributions, the desired sum
rules for the $f_i^T$ and $g_i^T$ form factors are obtained from the
following equations,
\bea
\label{nolabel}
f \Bigg\{ \Bigg( {m_2 \over m_1} -1 \Bigg) f_1^T + f_2^T \Bigg\}
e^{-m_1^2/M^2} \es \Pi_1^B~, \nnb \\
2 m_2 f e^{-m_1^2/M^2} f_2^T \es \Pi_2^B~, \nnb \\
f \Bigg\{ \Bigg( {m_2\over m_1} +1 \Bigg) g_1^T + g_2^T \Bigg\}
e^{-m_1^2/M^2} \es \Pi_3^B~, \nnb \\
- 2 m_2 f e^{-m_1^2/M^2} g_2^T \es \Pi_4^B~, 
\eea
where $\Pi_i^B$ are the Borel transformed coefficients of the
structures mentioned above, and $M^2$ is the Borel mass parameter.

The Borel transformation and continuum subtraction is performed with the help of the following master
formula.
\bea
\label{edb??}
&& \int dw {\rho(u,w) \over ( \Delta - m_Q^2)^n} = \nnb  \sum_{n=1}^{\infty} \Bigg\{ (-1)^n \int_0^{w_0} dw {e^{-s/M^2} 
\over  (n-1)!(M^2)^{n-1}} I_n  \nnb \\
\ek \Bigg[ {(-1)^{n-1} \over (n-1)!} e^{-s/M^2} \sum_{j=1}^{n-1}
 {1 \over (M^2)^{n-j-1}} {1\over s^\prime} \Bigg( {d\over dw} 
 {1 \over s^\prime}\Bigg)^{j-1} I_n \Bigg]_{w=w_0}\Bigg\}~,\nnb
\eea
where 
\bea
\label{edb??}
s \es {m_Q^2 - \bar{u} w (\bar{u} w - m_{2}) - \left(1 -
{\ds{\bar{u} w
    \over  m_{2}}}\right) q^2 \over {\ds {\bar{u} w \over  m_{2}}}}~, \nnb \\
I_n \es \rho(u,w) \over \left( {\ds {\bar{u} w \over m_{2}} } \right)^n
\eea

Note that $w=w_{0}$ is the solutions of the equations $s=s_{th}$
(and also $s=4 m_Q^2$), $s^\prime = {\ds {ds\over dw}}$, and the differential
operator is defined as,
\bea
\label{nolabel}
\Bigg({d\over dw} {1\over
s^\prime}\Bigg)^{j-1} I_n \to 
{d\over dw} \Bigg[ {d\over dw} {1\over
s^\prime} \cdots I_n \Bigg]~,
\eea
where $\cdots$ means the operation should be repeated $j-1$ times.
\section{Numerical Analysis}
\label{sec:3}
The primary aim of this section is to determine the $q^2$ dependence of the
form factors $f_1^T$, $f_2^T$, $g_1^T$, and $g_2^T$, whose LCSR are derived
in the previous section. Then we estimate the branching ratios of the
$\Xi_{QQ} \to \Sigma_Q \ell^+ \ell^-$ decays.

The LCSR of the form factors contain numerous input parameters. In further
numerical analysis, we choose the masses of the heavy quarks in the 
$\overline{MS}$ scheme, i.e., $\overline{m_c}{(\overline{m_c})} = (1.28 \pm 0.03)~\rm{GeV}$, and
$\overline{m_b}(\overline{m_b}) = (4.18 \pm 0.03)~\rm{GeV}$ \cite{PhysRevD.98.030001}. The masses, lifetime
and decay constants $f$ of the doubly heavy baryons are given in
Table~\ref{tab:1} (see also \cite{2014PhRvD..90i4007K,Aliev:2012ru,Shah:2016vmd,Shah:2017liu,Kiselev:2001fw}).

\begin{table}[h]
\centering
\renewcommand{\arraystretch}{1.3}
\setlength{\tabcolsep}{7pt}
\begin{tabular}{cccc}
\toprule
  Baryons  &        Mass $(\rm{GeV})$     & Life time $(fs)$ &    $f~(\rm{GeV})^3$~\cite{Hu:2017dzi} \\
  \midrule
$\Xi_{cc}^{++}$   &  3.621 \cite{LHCb:2017iph}  & 256 {\cite{LHCb:2018zpl}}     & $0.109 \pm 0.021$ \\
$\Xi_{cc}^{+}$    &  3.621 \cite{LHCb:2017iph}  &  45 {\cite{Cheng:2018mwu}}    & $0.109 \pm 0.021$ \\
$\Xi_{bb}^{0}$    & 10.143 \cite{Brown:2014ena} & 370 {\cite{Karliner:2017qjm}} & $0.281 \pm 0.071$ \\
$  \Xi_{bb}^{-}$  & 10.143 \cite{Brown:2014ena} & 370 {\cite{Karliner:2017qjm}} & $0.281 \pm 0.071$ \\
  \bottomrule
\end{tabular}
\caption{The mass, decay constants and lifetimes of the doubly heavy
$\Xi_{QQ}$ baryons.}
\label{tab:1}
\end{table}   

The mass and decay constants of the $\Sigma_Q$ baryon are chosen as
$m_{\Sigma_c} = 2.454~\rm{GeV}$, $m_{\Sigma_b} = 5.814~\rm{GeV}$, and $f^{(1)} =
f^{(2)} = 0.38$ \cite{Groote:1996em}

In addition to these input values, two extra auxiliary parameters, continuum threshold $s_{th}$, and the Borel mass parameter $M^2$, appear in the LCSR method.  These parameters are determined with  the following criteria. The working
region of $M^2$ is determined by requiring that the power corrections and
continuum contributions both should be suppressed compared to the leading twist-2 contribution. The
continuum threshold $s_{th}$ is determined so that the mass sum
rule reproduces the experimentally measured value of mass to within $\pm 5 \%$
accuracy.

Based on these conditions imposed by the LCSR method, we obtain the following 
working regions of the parameters $s_{th}$ and $M^2$ for the transitions
under consideration, i.e., $s_{th}=(16\pm 1)~\rm{GeV}^2$, $M^2 = (10\pm 2)~\rm{GeV}^2$ 
for the $\Xi_{cc} \to \Sigma_c$ transition, and $s_{th}=(112\pm 2)~\rm{GeV}^2$,
$M^2 = (20\pm 2)~\rm{GeV}^2$
for the $\Xi_{bb} \to \Sigma_b$ transition, respectively.
It should be emphasized here that these working regions are more or less
in the same range as those determined for the transitions induced by the
charged current \cite{Hu:2019bqj}.

It should be reminded that LCSR predictions are reliable in the low-energy region. Our
calculations show that the sum rule for the form factors is
meaningful in the domains $q^2 \le 0.5~\rm{GeV}^2$ for the
$\Xi_{cc} \to \Sigma_c$ transition and $q^2 \le 10~\rm{GeV}^2$ for the 
$\Xi_{bb} \to \Sigma_b$ transition, respectively.

Having determined the working regions for the QCD results of the form factors, we can extend the LCSR predictions  to the entire physical region. For this goal, we extrapolate these form factors 
to the physical region in such a way that in the region where LCSR is reliable, the result of the fit formula and the result LCSR method coincide with each other. Our analysis shows that the best-fit formula which satisfies the required restrictions is given as,
\bea
\label{nolabel}
F_i^T(q^2) = {f_i^T(0) \over 1 - {\ds{q^2 \over m_{fit}^2}} + \delta \Bigg({\ds{q^2 \over
m_{fit}^2}}\Bigg)^2}~,
\eea
where the values of the fit parameters $f_i^{T}(0)$, $m_{fit}$, and $\delta$ are presented in Table~\ref{tab:2}.

\begin{table}[h]
\centering
\renewcommand{\arraystretch}{1.3}
\setlength{\tabcolsep}{7pt}
\begin{tabular}{lccc}
\toprule
  Form factors (c-sector) & $f(0)$           & $m_{fit}$ & $\delta$ \\
  \midrule
$f_1^T(q^2)$              & $3.05 \pm 0.7$   & 1.69      & 0.43     \\
$f_2^T(q^2)$              & $0.49 \pm 0.1$   & 1.15      & 4.21     \\
$g_1^T(q^2)$              & $1.44 \pm 0.26$  & 1.43      & 0.42     \\
$ g_2^T(q^2)$             & $0.53 \pm 0.13$  & ---       & ---      \\
  \midrule
  Form factors (b-sector) &                  &           &          \\
  \midrule
$f_1^T(q^2)$              & $-1.72 \pm 0.35$   & 3.60      & 0.31     \\
$f_2^T(q^2)$              & $0.32  \pm 0.06$ & 3.49      & 0.75     \\
$g_1^T(q^2)$              & $0.21  \pm 0.04$ & ---       & ---      \\
$ g_2^T(q^2)$             & $0.32  \pm 0.06$ & 2.93      & 0.42     \\
  \bottomrule
\end{tabular}
\caption{The values of the fit parameters for the form factors of the
  $\Xi_{QQ} \to \Sigma_Q$ transition.}
\label{tab:2}
\end{table}

\begin{table}[h]
  \centering
\renewcommand{\arraystretch}{1.3}
\setlength{\tabcolsep}{7pt}
\begin{tabular}{lccc}
\toprule
  Branching Ratios                       & Ours                   & \cite{Xing:2018lre} & \cite{Hu:2020mxk}  \\
  \midrule
  $\Xi_{cc}^{++} \to \Sigma_c^{++} \ell^+ \ell^- $    & $1.62 (1 \pm 0.45)\times 10^{-13}$   &    ---       &    ---      \\
  \midrule
$\Xi_{bb}^0    \to \Sigma_b^0    e^+ e^-    $   & $2.51 (1 \pm 0.35) \times 10^{-8}$  &  $9.00 \times 10^{-9}$         &  $5.91 \times 10^{-9}$        \\
$\Xi_{bb}^0    \to \Sigma_b^0    \mu^+\mu^-  $  & $1.73 (1 \pm 0.30) \times 10^{-8}$  &  $7.94 \times 10^{-9}$         &  $4.89 \times 10^{-9}$        \\
$\Xi_{bb}^0    \to \Sigma_b^0    \tau^+\tau^- $ & $1.50 (1 \pm 0.31) \times 10^{-9}$  &  $1.18 \times 10^{-9}$         &  $1.91 \times 10^{-10}$        \\
\bottomrule
\end{tabular}
\caption{The branching ratios of the $\Xi_{QQ} \to \Sigma_Q \ell^+ \ell^-$ decays}
\label{tab:3}
\end{table}

The errors presented in the values of $f_i^{T}(q^2 = 0)$ point are due to the uncertainties in the mass of the heavy quark, Borel mass parameter, continuum threshold $s_{th}$, as well as from the input parameters appearing
in the DAs of the $\Sigma_b$ baryon. 

Having the results for the form factors, we
now proceed to calculate the corresponding branching ratios of the $\Xi_{cc}^{++} \to \Sigma_c^{++}
\ell^+\ell^-$ and $\Xi_{bb}^0 \to \Sigma_b^0 \ell^+\ell^-$ decays. Using the
definition of the matrix element of the $\Xi_{bb} \to \Sigma_b \ell^+ \ell^-$  the decay width is obtained as
\bea
\label{nolabel}
{d\Gamma (s) \over ds} \es {G_F^2 \alpha^2 m_1 \over 4096 \pi^5} |V_{tb} V_{td}^*|^2 v
\sqrt{\lambda(1,r,s)} \Bigg[T_1(s) + {1\over 3} T_2(s) \Bigg]~ \nnb
\eea
where $v = \sqrt{1 -{\ds{4 m_\ell^2
\over q^2} } }$ is the lepton velocity, $\lambda(1,r,s) = 1 + r^2 +s^2 - 2 r
- 2 s - 2 r s$, $s = {\ds {q^2\over m_1^2}}$, and  $r = {\ds {m_2^2\over
m_1^2}}$.  The lengthy expressions $T_1(s)$ and $T_2(s)$ can be found in~\cite{Aliev:2022maw}

Performing integration over the parameter $s$ in the domain ${\ds { 4m_l^2 \over m_1^2}}
\le s \le (1-\sqrt{r})$, and using the lifetimes of $\Xi_{cc}^{++}$,
$\Xi_c^+$, and $\Xi_{bb}^0$, we calculate the branching ratios of the
$\Xi_{QQ} \to \Sigma_Q$ decays, whose numerical results are all presented in
Table~\ref{tab:3}. For comparison, we also present the corresponding
branching ratios predicted by the Light Front approach \cite{Xing:2018lre,Ke:2019lcf}.
We observe from this comparison that our prediction for the $\Xi_{QQ} \to
\Sigma_Q \ell^+ \ell^-~(\ell = e,\mu)$ transition is larger than the predictions of the Light Front approach.
Considering the results summarized in Table~\ref{tab:3}, one can conclude that
the branching ratios of the 
$\Xi_{bb} \to \Sigma_b \ell^+ \ell^-$ decays could be
measured in future experiments at LHCb, while the measurement 
of the branching ratios of the $\Xi_{cc} \to \Sigma_c \ell^+ \ell^-$
decays presents quite a complex problem.
\section{Conclusion}
\label{sec:conclusion}
The decays induced by the flavor-changing neutral currents $b \to d$ and $ c \to u$ of the doubly heavy baryons are studied in the framework of the Light Cone Sum Rules method. We derive the LCSR of the form factors induced by the tensor
current. Using the results of the form factors obtained, we estimated the
corresponding branching ratios.
We found out that the branching ratios for $\Xi_{bb} \to \Sigma_b \ell^+ \ell^-$ ($\ell = e, \mu$) is at the order of $\sim 10^{-8}$. Moreover, for $\Xi_{cc} \to \Sigma_c \ell^+ \ell^-$ decay, the branching ratios are much smaller and at the order of $\sim 10^{-13}$. The relatively large value of the branching ratio $\Xi_{bb} \to \Sigma_b \ell^+ \ell^-$ indicates the possibility of being observed in future experiments at LHCb. 

Future improvements for the DA's of the $\Sigma_Q$ baryon and the inclusion of the gluon radiative corrections to the correlation function could pave the way to more
accurate sum rules and numerical predictions.


\appendix*
\section{The expression of the invariant functions $\rho_n^{(i)} (u,w)$}
\label{app:formulas}

\bea
\label{Eq17}
\rho_1^{(1)} \es f^{(2)} w \, \psi_3^{(s)}(u,w)~, \nnb \\
\rho_2^{(1)} \es \bar{u} \Bigg\{\Big[f^{(1)} m_Q \, \widehat{\psi}_4 (u,w)\Big] +
2 f^{(2)} \Big[ \bar{u} w - (q\mcdot v) \Big]
     \, \widehat{\psi}_3^{(\sigma)}(u,w) \nnb \\
\ek f^{(1)} \Big[m_Q - 4 (q\mcdot v)\Big] \, \widehat{\psi}_2 (u,w) -
    4 f^{(2)} \bar{u} \Big[w \, \widehat{\psi}_3^{(s)}(u,w) + 
2 \; {\widehat{\!\widehat\psi}}{}_3^{(\sigma)}(u,w) \Big] \Bigg\}~, \nnb \\
\rho_3^{(1)} \es 8 f^{(2)} \bar{u}^2 \Big[q^2 - \bar{u}^2 w^2 - 2 m_Q \, (q\mcdot
v)\Big]
   \; {\widehat{\!\widehat\psi}}{}_3^{(\sigma)}(u,w)~,\nnb \\
\rho_1^{(2)} \es  2 f^{(2)} \bar{u} \Big[w^2 \, \psi_3^{(s)}(u,w) + \, 
\widehat{\psi}_3^{(\sigma)}(u,w)\Big]~,\nnb \\
\rho_2^{(2)} \es  - 4 \bar{u} \Bigg\{f^{(1)} q^2 \, \widehat{\psi}_2(u,w) +
    f^{(2)} \bar{u} \Big[w (q\mcdot v) \, \widehat{\psi}_3^{(\sigma)}(u,w) \nnb \\
\ar  2 m_Q \; {\widehat{\!\widehat\psi}}{}_3^{(\sigma)}(u,w) + \bar{u} w \Big(w \widehat{\psi}_3^{(s)}(u,w) +
        4 \; {\widehat{\!\widehat\psi}}{}_3^{(\sigma)}(u,w)\Big)\Big]\Bigg\}~, \nnb \\
\rho_3^{(2)} \es 16 f^{(2)} \bar{u}^2 \Big[q^2  (m_Q + \bar{u} w) +
   \bar{u}^2 w^2 (q\mcdot v) \Big]\; {\widehat{\!\widehat\psi}}{}_3^{(\sigma)}(u,w)~,
\nnb \\
\rho_1^{(3)} \es -f^{(2)} w \, \psi_3^{(s)}(u,w)~, \nnb \\
\rho_2^{(3)} \es  \bar{u} \Bigg\{f^{(1)} m_Q \, \widehat{\psi}_4 (u,w) - 
2 f^{(2)} \Big[\bar{u} w - (q\mcdot v)\Big]
     \, \widehat{\psi}_3^{(\sigma)}(u,w) \nnb \\
\ek f^{(1)} \Big[m_Q - 4 (q\mcdot v)\Big] \,\widehat{\psi}_2(u,w) 
+ 4 f^{(2)} \bar{u} \Big[w \, \widehat{\psi}_3^{(s)}(u,w) + 
2 \; {\widehat{\!\widehat\psi}}{}_3^{(\sigma)}(u,w) \Big] \Bigg\} ~, \nnb \\
\rho_3^{(3)} \es  - 8 f^{(2)} \bar{u}^2 \Big[q^2 - 
\bar{u}^2 w^2 - 2 m_Q (q\mcdot v) \Big]
   \; {\widehat{\!\widehat\psi}}{}_3^{(\sigma)}(u,w)~, \nnb \\
\rho_1^{(4)} \es - 2 f^{(2)} \bar{u} \Big[w^2 \psi_3^{(s)}(u,w) + 
\, \widehat{\psi}_3^{(\sigma)}(u,w) \Big]~, \nnb \\
\rho_2^{(4)} \es - 4 \bar{u} \Bigg\{\Big[f^{(1)} q^2 \, \widehat{\psi}_2 (u,w)\Big] -
    f^{(2)} \bar{u} \Big[w (q\mcdot v)  \, \widehat{\psi}_3^{(\sigma)}(u,w) \nnb \\
\ar 2 m_Q \; {\widehat{\!\widehat\psi}}{}_3^{(\sigma)}(u,w) + \bar{u} w
\Big(w \, \widehat{\psi}_3^{(s)}(u,w) +
        4 \; {\widehat{\!\widehat\psi}}{}_3^{(\sigma)}(u,w) \Big) \Big]
\Bigg\}~, \nnb \\
\rho_3^{(4)} \es - 16 f^{(2)} \bar{u}^2 \Big[q^2  (m_Q + \bar{u} w ) +
    \bar{u}^2 w^2 (q\mcdot v) \Big] \; {\widehat{\!\widehat\psi}}{}_3^{(\sigma)}(u,w) ~. 
\eea

The functions $\widehat{\psi}(u,w)$ and ${\widehat{\!\widehat\psi}}{}(u,w)$
are defined as,

\bea
\label{nolabel}
\widehat{\psi}(u,w) \es \int_0^w d \tau^\prime \psi(u,\tau^\prime)
\tau^\prime~,\nnb \\
{\widehat{\!\widehat\psi}}(u,w) \es
\int_0^w d \tau^\prime \widehat{\psi}(u,\tau^\prime)~.\nnb
\eea

\bibliographystyle{utcaps_mod}
\bibliography{../all.bib}



\end{document}